# Thermal spin-crossover and temperature-dependent zero-field splitting in magnetic nanographene chains

Yan Wang[1], Alejandro Pérez Paz[2], Emil Viñas Boström[3,4], Xiaoxi Zhang[1], Juan Li[5], Reinhard Berger[6], Kun Liu[9], Ji Ma[9], Li Huang[1], Shixuan Du[1], Hong-jun Gao[1,8], Klaus Müllen[6], Akimitsu Narita[6,7], Xinliang Feng[9,10], Angel Rubio[2], CA Palma[1,11]

[1]Institute of Physics, Chinese Academy of Sciences, Beijing 100190, China
[2]Chemistry Department, United Arab Emirates University, P.O. Box 15551, Al Ain, United Arab Emirates
[3]Nano-Bio Spectroscopy Group, Departamento de Física de Materiales, Universidad del País Vasco, 20018 San Sebastian, Spain
[4]Max Planck Institute for the Structure and Dynamics of Matter, Luruper Chaussee 149, 22761 Hamburg, Germany
[5]Advanced Research Institute for Multidisciplinary Science, Beijing Institute of Technology, 100081 Beijing, China.
[6]Max Planck Institute for Polymer Research, Ackermannweg 10, 55128 Mainz, Germany
[7]Organic and Carbon Nanomaterials Unit, Okinawa Institute of Science and Technology Graduate University, 1919-1 Tancha, Onna-son, Kunigami-gun, Okinawa 904-0495, Japan
[8]School of Physical Sciences, University of Chinese Academy of Sciences, Beijing 100190, China
[9]Center for Advancing Electronics Dresden (cfaed) & Faculty of Chemistry and Food Chemistry, Technische Universität Dresden, D-01069 Dresden, Germany
[10]Max Planck Institute of Microstructure Physics, Weinberg 2, 06120 Halle, Germany
[11]Department of Physics& IRIS Adlershof, Humboldt-Universität zu Berlin, 12489 Berlin, Germany

**Abstract.** Nanographene-based magnetism at interfaces offers an avenue to designer quantum materials towards novel phases of matter and atomic-scale applications. Key to spintronics applications at the nanoscale is bistable spin-crossover which however remains to be demonstrated in nanographenes. Here we show that antiaromatic 1,4-disubstituted pyrazine-embedded nanographene derivatives, which promote magnetism through oxidation to a non-aromatic radical are prototypical models for the study of carbon-based thermal spin-crossover. Scanning tunneling spectroscopy studies reveal symmetric spin excitation signals which evolve at $T_c$ to a zero-energy peak, and are assigned to the transition of a $S = 3/2$ high-spin to a $S = 1/2$ low-spin state by density functional theory. At temperatures below and close to the spin-crossover $T_c$, the high-spin $S= 3/2$ excitations evidence pronouncedly different temperature-dependent excitation energies corresponding to a zero-field splitting in the Hubbard-Kanamori Hamiltonian. The discovery of thermal spin crossover and temperature-dependent zero-field splitting in carbon nanomaterials promises to accelerate quantum information, spintronics and thermometry at the atomic scale.

The fabrication of carbon nanomaterials at metal interfaces is an established strategy for studying magnetism at the nanoscale (*1-12*). Yet metal substrates can influence spin-bearing carbon centers and could be detrimental to their magnetic properties. Moreover, magnetic carbon centers, which are radical sites in terms of organic chemistry, are usually highly reactive, leading to additional integration challenges towards applications. Antiaromatic 1,4-disubstituted pyrazine centers, for example in diazahexabenzocoronene (diaza-HBC) synthesized from polycyclic azomethine ylides (PAMY (*13-17*)), offer a new paradigm to organic magnetism: Antiaromatic and thus unstable diaza-HBC is oxidized to form a stable magnetic radical cation, diaza-HBC$^{•+}$, on Au(111) due to the electron transfer, as opposed to unstable magnetic centers which are prone to the loss of the unpaired electrons or other passivation reactions. Diaza-HBC as the molecular magnet (*14, 18, 19*) has been recently reported to promote spin 1/2 Kondo-chains (*18*), potentially realizing Heisenberg spin 1/2 - chains (*20-26*). Therefore, diaza-HBC chains could offer an entry to study one dimensional spin-phase transitions between Kondo (*27-31*) and spin regimes, Kondo transition peak splitting (*32*) as well as spin-chain excitations at the thermodynamic limit. Critical to the understanding of spin excitations, is the study of temperature-dependent spin states in carbon magnets. For example, thermal spin-crossover (SCO) has been observed on metal-bearing molecular magnets at interfaces (*33, 34*), and mechanistically assigned to thermodynamic and structural (volumetric) changes (*35*). Specifically, thermal SCO, whereby a molecule can be switched between low-spin LS at low- and high-spin HS at high temperature, has been suggested to occur when the spin-pairing energy lies between the gap of a HS and LS state, with the enthalpy of the HS state being slightly higher, and smooth transitions attributed to volumetric effects(*36*). Similarly, volumetric changes and spin-spin dipole interactions have been proposed to account for the temperature-dependent zero-field splitting in diamond NV-centers (*37, 38*). Altogether, temperature-dependent, nanoscale studies of bistable (*39, 40*) carbon-based magnetism are instrumental for the understanding of spin transitions and the design of macroscopic applications based on robust physical principles at the atomic-scale.

Here we observe two different spin ground-states at different temperatures in diaza-HBC **3** with PAMY side groups and its dimer **4**, fabricated from the reaction of dimeric polycyclic aromatic azomethine ylide (diPAMY) precursor **1** (*41*) on Au(111) (**Figure 1a**). Zero-energy peaks and symmetric peaks in scanning tunneling spectroscopy (STS) are found for both **3** and **4** above 4 K, respectively, which have been recently explained by the formation of a S = 1/2 radical cation **3**$^{•+}$ (**Figure 1b**), S=1 diradical dication **4**$^{2•2+}$ and their Heisenberg spin-chains (*41, 42*). At temperatures close to 2 K, symmetric peaks evolve at negative and positive biases. For **3**$^{•+}$ and **4**$^{2•2+}$, temperature-dependent measurements during scanning tunneling spectroscopy identify the ZEP and symmetric signals as co-existing in an intermediate temperature regime between 2 K and 4 K (**Figure 1c**). Density functional theory (DFT) corroborates the surface oxidation of 3 to radical cation **3**$^{•+}$, predicting two accessible states, namely low-spin S = 1/2 and high-spin S = 3/2. A high-spin to low-spin spin cross-over is modelled by a Hubbard-Kanamori Hamiltonian model with a temperature (T)-dependent. zero-field splitting (ZFS) parameter D'(T) in the S = 3/2 ground state.

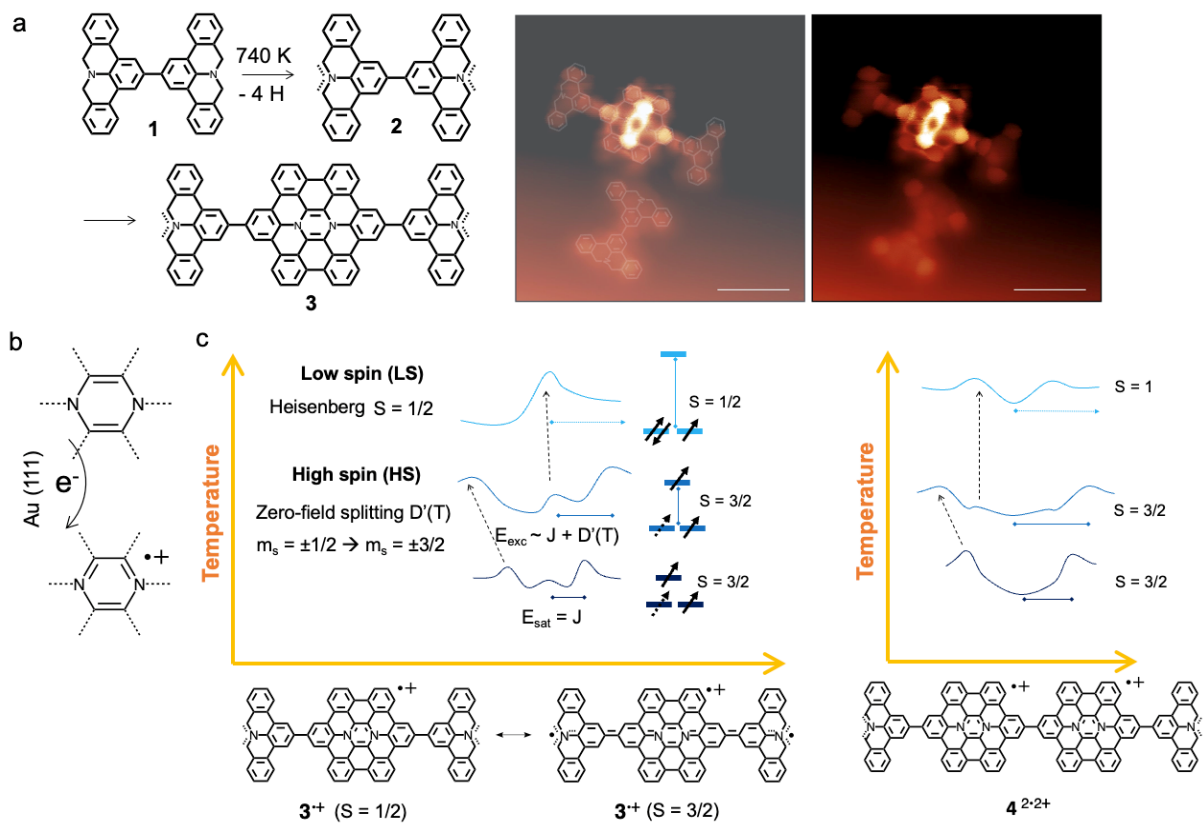

**Figure 1**. **a** Reaction of diPAMY precursor **1** for the formation of **3** (cf. Figure S1–S3). Scanning tunneling microscopy of **3** and the dehydrogenated precursor **2** obtained with a bond-resolving tip. $V_b$ = -50 mV, scale bar = 1 nm. The lower current contrast near at the periphery of **2** and **3** indicates that the CH$_2$ groups (precursor **1**) have been dehydrogenated to CH. **b** Mechanism of the antiaromatic radical cation formation of diazabenzene core of **3**$^{•+}$ through electron transfer **c** Schematics of a high-spin state S = 3/2, low-spin state S = 1/2 of **3**$^{•+}$ and a coexisting, temperature dependent region. A three orbital model describes the high-spin S = 3/2 excitation energy $E_{exc}$ governed by an exchange coupling J and temperature-dependent zero-field splitting D'(T) and leading to the spin-crossover to S = 1/2 at higher temperatures.

Scanning tunneling spectroscopy characteristics of **3**$^{•+}$ build at 4.3 K reveal a zero-energy peak at its central position (ZEP, **Figure 2a**). When cooling to 2 K, the ZEP almost completely disappears and the ZEP evolution is conspicuous when warming up the system, starting at 2.93 K. At the same time, from 2 K to 3 K, symmetric peaks are concomitantly observed below $E_{exc}$ = ± 9.6 mV, interpreted as inelastic spin excitations. The spin excitation energy $E_{exc}$ increases between 2.7 to 3.0 K, the excitations moving away from the Fermi level by more than 20 meV. Overall, the transition between the ZEP resonance and $E_{exc}$ signals are smooth, and evidence coexistence of ZEP and $E_{exc}$ states. These observations can be interpreted through the existence of a magnetic moment below 3 K which is strongly temperature dependent at lower temperatures. At the same time, the current micrographs between low and high temperatures show marked differences. Above 4 K, the diaza-HBC core and the PAMY side groups of **3$^{•+}$** can be resolved. The high-contrast current data (**Figure 2a**, right hand side) clearly identifies the positions of the benzene rings. Previous work (*17*) predicts that this PAMY structure alone has an open shell singlet ground state in the gas phase, with the triplet state higher in energy by >1 eV (25 kcal mol$^{-1}$).

Below 3 K, STM data of **3$^{•+}$** is consistently blurred (**Figure 2a**, left hand side), whereby the low temperature STM data appears up to 0.4 Å longer than the high temperature (**Figure S6**). Noteworthy, varying the distance between the tip and the sample during spectroscopy induces peak shifts for the high temperature ZEP and $E_{exc}$, of 0.02 and 0.03 meV, respectively. The orange star marks the initial measurement position, the blue marks the position after the tip was moved down by 0.05 nm, and the green marks the position after the tip was moved down by 0.1 nm relative to the initial position. At 2.5 K, as the tip approaches the core of **3$^{•+}$**, the intensity of the conductance increases corresponding to several fold increase of the current, while the symmetric peaks shift from $E_{exc}$ = ± 9.2 mV to $E_{exc}$ = ± 8.9 mV. The observations point towards a weak absolute tunneling current and tip-position dependence as compared to the effect of temperature.

The spectroscopy data of diaza-HBC dimer **4$^{2•2+}$** depicts symmetric peaks starting at ± 27 meV followed at low temperature by the evolution of two shallow, temperature-independent peaks between 3 and above 4 K at ± 33 meV (**Figure 3**). Compared to a ZEP of **3**, such temperature-independent, shallow peaks could be interpreted as an inverted Kondo similar to observations of Co atoms on Au(111) and Ce atoms on Ag(111). Alternatively, they could be interpreted as singlet-triplet excitation peaks stemming from the coupling of S=1/2 magnets (*41, 42*), **Figure S5**. Similar to the results of **3$^{•+}$** in **Figure 2**, the current topography of **4$^{2•2+}$** depicts blurred STM image data at low temperatures. Conversely, the temperature dependent sensibility observed for **4$^{2•2+}$** is weaker, and the symmetric peaks excitation energy, $E_{sat,}$ remains constant already below 2.9 K (as opposed to ~2.7 K for **3$^{•+}$**).

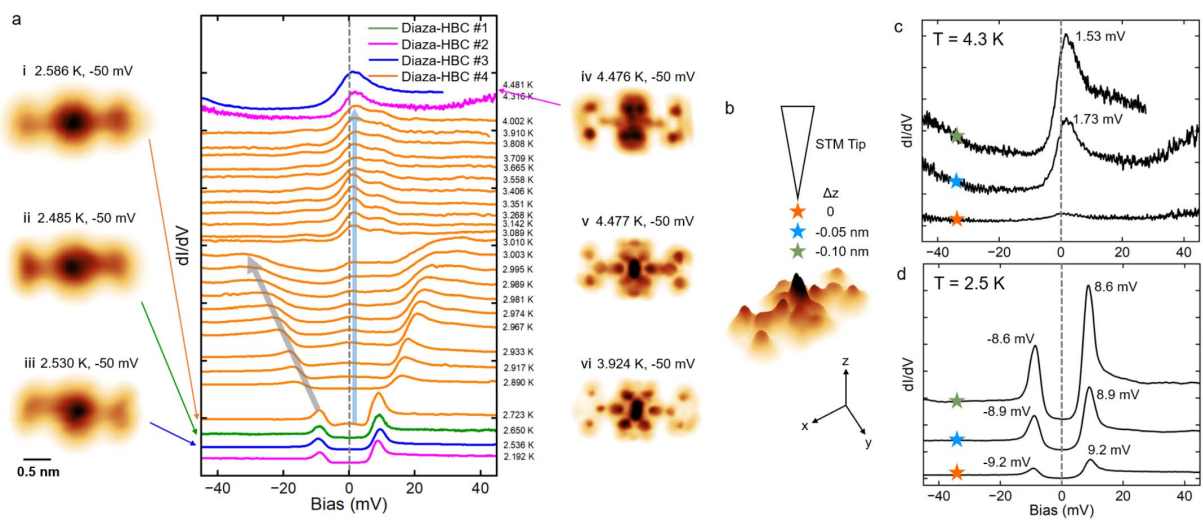

**Figure 2**. **a** STM current topography data and temperature dependent scanning tunneling spectroscopy of different three individual diaza-HBC **3**, i. $V_b$ = -50 mV, T = 2.586 K. ii. $V_b$ = -50 mV, T = 2.485 K. iii. $V_b$ = -50 mV, T = 2.530 K. iv. $V_b$ = -50 mV, T = 4.476 K. v. $V_b$ = -50 mV, T = 4.477 K. vi. $V_b$ = -50 mV, T = 3.924 K. **b** Topographic data depicting the tip position variation at **c** 4.3 K and **d** lower temperatures 2.5 K temperatures. $V_b$ = -50 mV, T = 4.477 K. **c** Height dependent scanning tunneling spectroscopy of **3** at 4.3 K. **d**. Height dependent scanning tunneling spectroscopy of **3** at 4.3 K.

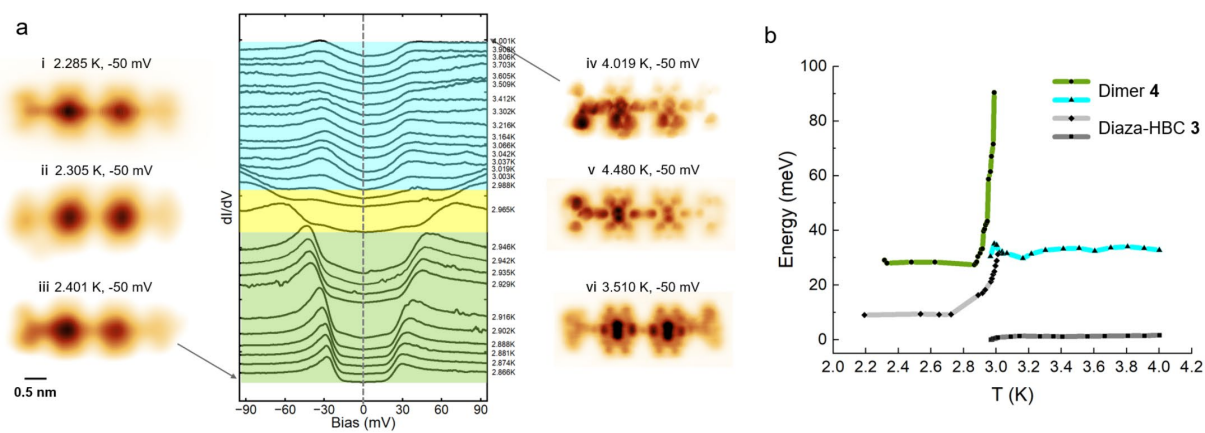

**Figure 3**. **a** STM current topography data and temperature dependent scanning tunnelling spectroscopy of three different **4$^{2•2+}$**, first dimer: i ($V_b$ = -50 mV, T = 2.285 K) and v ($V_b$ = -50 mV, T = 4.019 K), second dimer: ii ($V_b$ = -50 mV, T = 2.305 K) and vi ($V_b$ = -50 mV, T = 4.480 K), third: iii ($V_b$ = -50 mV, T = 2.401 K) and iv ($V_b$ = -50 mV, T = 3.510 K). **b** Scanning tunneling spectroscopy peaks vs temperature shows a step temperature dependence close to the transition temperature for both **3$^{•+}$** and its dimer **4$^{2•2+}$**.

Kondo resonances offer an explanation to a ZEP in radical cation **3$^{•+}$**, with the resonance resulting from the coupling between the spin of **3$^{•+}$** and the spin of the conduction electrons in the substrate. The shape of the Kondo resonance peak reflects the competition between tunneling electrons directly interacting with the localized spin and those indirectly interacting through spin-correlated surface electrons, largely depending on the coupling strength between the substrate and the adsorbate, determined by the entire adsorbate-substrate system (*43*). DFT calculations with an underlying gold predict a Bader charge transfer of 1.045 electrons donated to the Au slab, corroborating the magnetic moment of **3** as esteeming from a S = 1/2 doublet **3·+** on Au(111), **Figure 4a**. The gas-phase DFT identifies two degenerate orbitals product of the interaction with the PAMY-terminated zigzag edge of the **3·+** (**Figure 4b**): an occupied spin-down SOMO at -5.74 eV, quasi-degenerate with another orbital at -5.74 eV, and an unoccupied SUMO at -5.75 eV. The closest spin-up SOMO lies at -5.90 eV. Under this configuration, DFT predicts that the S = 3/2 quartet state as a high-spin state higher in energy by 0.98 eV. The quartet frontier orbital diagram is highly degenerate (**Figure 4b**) and is expected to be strongly affected by the substrate and spin polarized multi-reference quantum chemistry corrections, a level of theory which however is considered not accessible to date. Therefore, a stable S = 3/2 quartet state can be estimated from an effective model consisting of two degenerate orbitals, and a third level at a higher energy in agreement with the DFT. The eigenstates of the manifold can be described by a Hubbard-Kanamori Hamiltonian of the form

$$H_U = U \sum_\alpha \hat{n}_{\alpha\uparrow}\hat{n}_{\alpha\downarrow} + \sum_{\sigma\sigma',\alpha<\beta} (U' - J\delta_{\sigma\sigma'})\hat{n}_{\alpha\sigma}\hat{n}_{\beta\sigma'}$$
$$+ J \sum_{\alpha\neq\beta} (\hat{d}^\dagger_{\alpha\uparrow}\hat{d}^\dagger_{\alpha\downarrow}\hat{d}_{\beta\downarrow}\hat{d}_{\beta\uparrow} - \hat{d}^\dagger_{\alpha\uparrow}\hat{d}_{\alpha\downarrow}\hat{d}^\dagger_{\beta\downarrow}\hat{d}_{\beta\uparrow}), \qquad (1)$$

where U is a local Coulomb interaction (the Hubbard interaction), J is an exchange coupling (the Hund's coupling) favoring high-spin states, and U' = U − 2J. With this Hamiltonian, the single-hole states correspond to three-electron states, and have spin either S = 1/2 or S = 3/2. Introducing the total charge *N*, spin *S* and orbital angular *L* momentum operators(*44*), denoted by:

$$H_U = (U - 3J)\frac{\hat{N}(\hat{N}-1)}{2} - 2J\hat{S} - \frac{J}{2}\hat{L} + \frac{5J}{2}(\hat{N}-1). \qquad (2)$$

If the orbital magnetic moment is quenched (i.e., that L = 0), it leads to the energies $E_{S=1/2}$ = 3U − 5J and $E_{S=3/2}$ = 3U − 7J as long as the molecular orbitals are degenerate. In addition to the symmetric interactions, the observations can be modeled by a third orbital at an energy D above the other two (see Fig. 1), and that there is a zero-energy splitting D' between the S = 3/2 and S = 1/2 associated with orbital reconfigurations (structural volume changes or phonons). We then find an S = 3/2 state with energy $E_0$ = 3U − 7J + D − D', and two S = 1/2 states with energies $E_1$ = 3U − 5J and $E_2$ = 3U − 5J + D. In the former S = 1/2 state, all three electrons occupy the lower orbitals, while in the later state one electron is in the higher orbital.

For $\Delta E_1 = E_1 − E_0 = 2J − D > 0$ the ground state is the high-spin S = 3/2 state, while for $\Delta E_1 < 0$ the ground state is the low-spin S = 1/2 state (**Figure 4c**). With a temperature dependent parameter D = D(T), there is therefore a thermal SOC spin cross-over at the temperature where 2J / D(T) = 1 from a high-spin to a low-spin ground state.

At low temperatures, when the ground state is the high-spin S = 3/2 state, a single spin-flip in the lower states leads to an excited S = 1/2 state separated from the ground state by an energy $\Delta E_{exc}$ = 2J + D'(T ). Thus, if the energy D'(T) is assumed to be a linearly increasing function of temperature, there will be a linearly dispersing excitation peak $\Delta E_{exc}$(T) = 2J + D'(T ), (**Figure 4d**) whereas the spectroscopy data in **Figure 3b** appears to show slightly different dispersions of the excitation peak at the transition energy for **3•+**.

To model dimer **4 $^{2•2+}$**, we add to the local Hamiltonian an exchange coupling between the magnetic moments on each molecule. This interaction is given by:

$$H_K = K\mathbf{S}_1 \cdot \mathbf{S}_2 = \frac{1}{2}(\mathbf{S}^2 − \mathbf{S}_1^2 − \mathbf{S}_2^2). \qquad (3)$$

The eigenstates of this Hamiltonian are multiplets with spin length S ∈ S1 + S2, S1 + S2 −1,..., |S1 − S2| and energies:

$$E_S = \frac{K}{2}[S(S+1) − S_1(S_1+1) − S_2(S_2+1)]. \qquad (4)$$

In the high temperature state with local moments $S_1$ = 1/2, and assuming K > 0 (*42*) the ground state of **4$^{2•2+}$** is a singlet with energy $E_{singlet}$ = 3U − 5J − 3K/4 with an excited triplet state at energy $E_{triplet}$ = 3U − 5J + K/4. Therefore, for dimer **4 $^{2•2+}$**, we do not expect a Kondo peak at high temperature but rather two symmetric excitation peaks at ± K. In the low temperature regime, the excitation peaks are shifted outwards by an energy 3K/2. Fitting the model to the data of **3** and **4** (**Figure 3**) leads to a temperature dependence of the zero-energy splitting D'(T) of 10 mV K$^{-1}$.

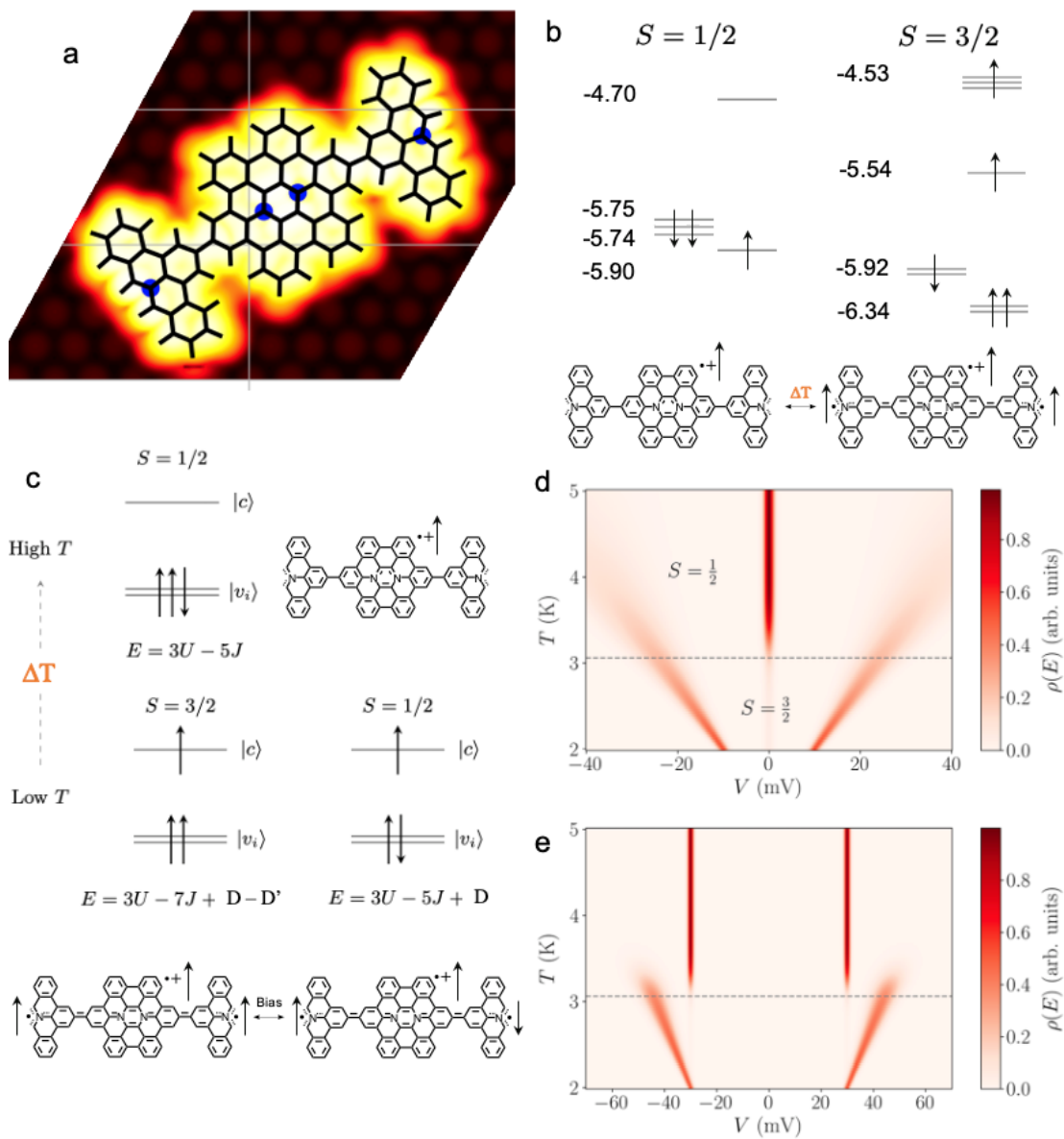

**Figure 4**. **a** DFT STM simulation of **3** leading to identification of a **3•+** species on an Au(111) slab. **b** Gas-phase orbital diagram for **3•+** in doublet and quartet state showing markedly different orbital energies. **c** Schematic depiction of the Hubbard-Kanamori Hamiltonian low and high temperature spin states, and the model for temperature induced transition from a high-spin (S = 3/2) to a low-spin (S = 1/2) ground state including a zero-energy splitting D'(T) for the S = 3/2 high-spin state. **d** Fitting of the Hubbard-Kanamori Hamiltonian with D'(T) = 10 mV K$^{-1}$ and a local exchange coupling J for **3•+** and **4 2•2+** depicting the local density of states ρ(E − eV ) as a function of bias voltage V and temperature T.

The microscopic thermal dependence mechanism of the spin-crossover energy parameter D(T) and the zero-energy splitting parameter D'(T) is substantiated by STM structural differences between low and high temperature, corresponding to DFT geometry changes leading to markedly different orbital energies between the S = 1/2 and S = 3/2 states. Our work unveils a mechanism for which carbon nanomaterials undergo thermal spin-crossover and zero-energy splitting.

## Methods

**Sample preparation and STM/STS measurements.** STM measurements was performed with a CreaTec 1.2 K-STM operating at base pressure below 1.0E-9 mbar. Au(111)/mica surfaces were prepared by several cycles of $Ar^+$ sputtering and subsequent annealing to 723 K. DiPAMY precursor **1** was prepared as reported in reference (*41*) and contained in a quartz crucible and sublimed from a CreaTec evaporator at 423 K onto Au(111) surface held at ~200 K, subsequently annealed at 373 K to remove residual impurities and over 740 K to induce reactions to form **3**. STM images were recorded in constant-current mode by Pt/Ir tip. STS was measured with a lock-in amplifier (f = 1 kHz). During temperature-dependent scanning tunneling spectroscopy (STS) measurements, we first measure STS data and then pumped the 1K pot from ~1013 mbar to 1 mbar, which simultaneously cooled the sample from 4.5 K to 2.1 K for ~1 h. After stopping the pumping, we allowed the sample to warm up and performed temperature-dependent STS measurements.

**Simulations.** Only **3** was considered for simulation in this work (**Figure 1**). The slab calculations were carried out with the program GPAW (version 22.8.0) in conjunction with the code ASE(*45*). The program(*46*) was run in the LCAO mode(*47*) using 'dzp' quality basis set. In all DFT calculations, the Grimme's "D3" van der Waals corrections were included, which are essential to get the right adsorption geometries and energetics. The PBE exchange-correlation functional was used. The relaxed dimer physisorbs flat on the surface with a mean vertical separation of about 3.06 Å from the top layer of Au(111). Given the large size of our system (486 atoms) and the unit cell (side length ~ 31 Å) a frozen 3-layer Au slab (363 gold atoms in total) with Cartesian coordinates derived from the experimental lattice constant of pure gold (4.0786 Å) was employed. The Gamma point was used for the reciprocal space sampling. The smearing was 0.1 eV and the convergence threshold was set to 1E-5 in the density for the SCF cycle. A Bader charge transfer of 1.045 $e^-$ from the adsorbate to the Au(111) surface was calculated. This prompts us to investigate the molecule mono-cation in gas phase as a simplified model of what might happen on the surface.

All gas phase spin-polarized calculations on the mono-cation were performed with the program ORCA(*48*) (version 5.0.3) of Frank Neese and co-workers. The DFT level of theory was Grimme's $r^2$SCAN-3c functional(*49*), which provides an excellent trade-off between accuracy and computational effort as it corrects for the common deficiencies of ordinary GGA functionals. The geometry of the doublet and quartet spin states were optimized. The wavefunction analysis was carried out with the program Multiwfn(*50*).

**Acknowledgements**

We acknowledge the financial support from the National Natural Science Foundation of China (No. 61888102, 61622116, 11974403, 12304238 and 62488201), the Strategic Priority Research Program of Chinese Academy of Sciences (No. XDB33000000 and XDB33030000), the German Research Foundation (DFG) with EnhanceNano (No. 391979941), Matters of Activity (390648296) the CAS Frontier Sciences and Education grant (No. QYZDBSSW-SLH038), the China Postdoctoral Science Foundation (2023M733721, 2024T170991), EVB acknowledges funding from the European Union's Horizon Europe research and innovation programme under the Marie Sklodowska-Curie grant agreement No 101106809. Joaquín Fernandez-Rossier and José Lado are gratefully acknowledged for the interpretation of the spectroscopy as a potential structural change and temperature-dependent distortion. Sheng Meng, Hongming Weng and Heng Fan are likewise acknowledged for discussions related to spin-chains. We thank Oliver Benson and Masazumi Fujiwara for discussions regarding temperature-dependent zero energy splitting. Nian Lin is acknowledged for a critical feedback on the manuscript.

**Author information**

**Contributions**

A.R., C.-A.P., and X.F. supervised the work; E.V.B. and A.R. developed the theoretical model and A.P.P. and A.R. the all-atomistic calculations; Y.W., X.Z., J.L. and C.-A.P. performed STM experiments; R.B., K.M., designed and J.M. and K.L. synthesized the molecules; L. H., S.X. D, H.-J.G., H. W. contributed to the discussion and E.V.B., A.N. to the interpretation. C.-A.P. conceived the project and designed the studies. The manuscript was written with contributions from all authors.

Supplementary information:

# Thermal spin-crossover and temperature-dependent zero-field splitting in magnetic nanographene chains

**Preparations and measurements**. diPAMY precursor was deposited onto the Au(111) surface, the sample was annealed at 400 K to remove residual impurities. The molecular structure on the substrate surface was characterized using LT-STM. diPAMY molecules exhibit a planar polycyclic aromatic structure with nitrogen atoms doped to modulate their electronic and structural properties (**Figure S1**). **Figure S1a** shows the top and side views of the diPAMY molecular structure model, with blue atoms representing the doped nitrogen atoms. The STM topographic image in **Figure S1b** shows individual diPAMY molecules adsorbed on the Au(111) surface, mostly distributed between the "herringbone" patterns of the Au(111) substrate. The results indicate that the diPAMY molecules adopt a flat-lying adsorption geometry on the surface, with single molecules appearing "H"-shaped, consistent with the molecular model. This symmetry and flat-lying configuration maximize van der Waals interactions between the molecules and the substrate, maintaining the most stable state. Furthermore, on the sample surface without high temperature annealing, we observed a lot of individual diPAMY molecules, suggesting that the molecules prefer to adsorb independently on the substrate surface. **Figures S1c-d** provide magnified models and STM topographic images of diPAMY molecules on the Au(111) surface, showing the consistency between the experimental results and the chemical structure. The apparent height of a single diPAMY molecule was measured to be 0.15 nm, corresponding to the side view of the model. Based on these experimental results, diPAMY molecules have adsorbed on the gold substrate surface, and we successfully prepared a submonolayer structure of HBC molecules. To prepare the diPAMY molecular chain structure, we annealed the submonolayer diPAMY samples at 740 K for several hours, inducing cyclodehydrogenation reactions of the surface molecules and successfully synthesizing diPAMY molecular chains. Our experiments revealed that the coverage rate of the molecules significantly impacts the yield of the final product. This is because the diffusion of surface molecules is crucial for the polymerization reaction, and an appropriate coverage rate is essential for free diffusion after dehydrogenation. Excessive coverage restricts the free diffusion of molecules, while insufficient coverage reduces the reaction probability between molecules, leading to lower polymer yield. Thus, during the deposition of precursor molecules, we prepared a submonolayer structure of diPAMY molecules, which we identified as the optimal coverage rate after multiple experiments. After annealing the diPAMY samples at 740 K, we studied the sample surface using STM. To observe the changes in the molecular configuration on the surface post-annealing, we performed extensive large-scale scans of the sample surface. The STM topographic images in **Figures S2a-d** sequentially show the sample surface at different sizes: 80 × 80 nm, 60 × 60 nm, 40 × 40 nm, and 20 × 20 nm. We identified three main molecular configurations on the sample surface: non-reactive diPAMY monomers, which account for approximately 30% of the surface; irregular molecular clusters, marked by white dashed lines in **Figure S2c**, formed by surface chemical reactions without forming regular ordered nanostructures and containing numerous defect structures, accounting for about 60% of the surface; and diPAMY molecular chains, dimers 4, trimers, and tetramers, marked by green, blue, and orange arrows in the STM topographic images, formed through cyclodehydrogenation reactions between molecules, and accounting for the smallest proportion, about 10% of the total. In this work, the diPAMY molecular chains are our desired final product. We counted the number of molecular chains of different lengths on a sample surface area of 110 × 110 nm, as shown in **Figure S3**: diaza-HBC had the highest yield with 27 counts, representing 87% of the total diPAMY molecular chains; trimers had the second highest yield with 3 counts, representing 10%; and tetramers had the lowest yield with only 1 count, representing 3%. **Figure S3** also shows the chemical structures corresponding to the molecular chains, with red frames marking the positions of cyclodehydrogenation reactions between molecules. We found that 740 K annealing can form well-defined diPAMY molecular chains on the sample surface, with a final product yield of about 10%, with single diaza-HBC being the most numerous, followed by trimers, and tetramers being the least numerous.

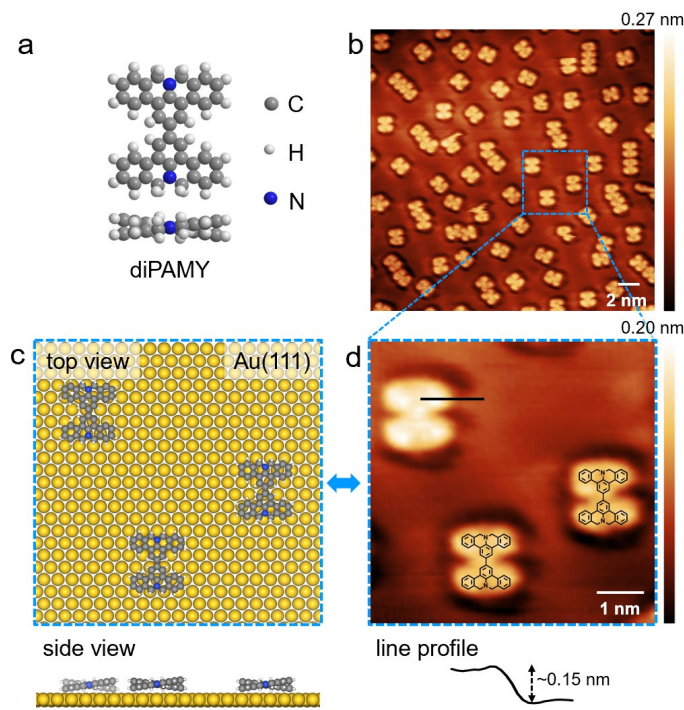

**Figure S1.** Morphological characterization of submonolayer diPAMY molecules on the Au(111) substrate. **a.** Top and side views of the diPAMY molecular structure model. **b.** STM topographic image of the submonolayer structure of diPAMY molecules on the Au(111) surface. Scanning parameters: 27 × 27 nm$^2$, $V_b$ = 20 mV, $I_t$ = 15 pA, $T_{STM}$ = 4.5 K. **c.** Magnified top and side views of the diPAMY molecular model on the Au(111) surface. **d.** Magnified STM topographic image of diPAMY molecules on the Au(111) surface and the height profile of a single molecule, with the chemical structure of diPAMY molecules superimposed on the topographic image. Scanning parameters: 6 × 6 nm$^2$, $V_b$ = 20 mV, $I_t$ = 15 pA, $T_{STM}$ = 4.5 K.

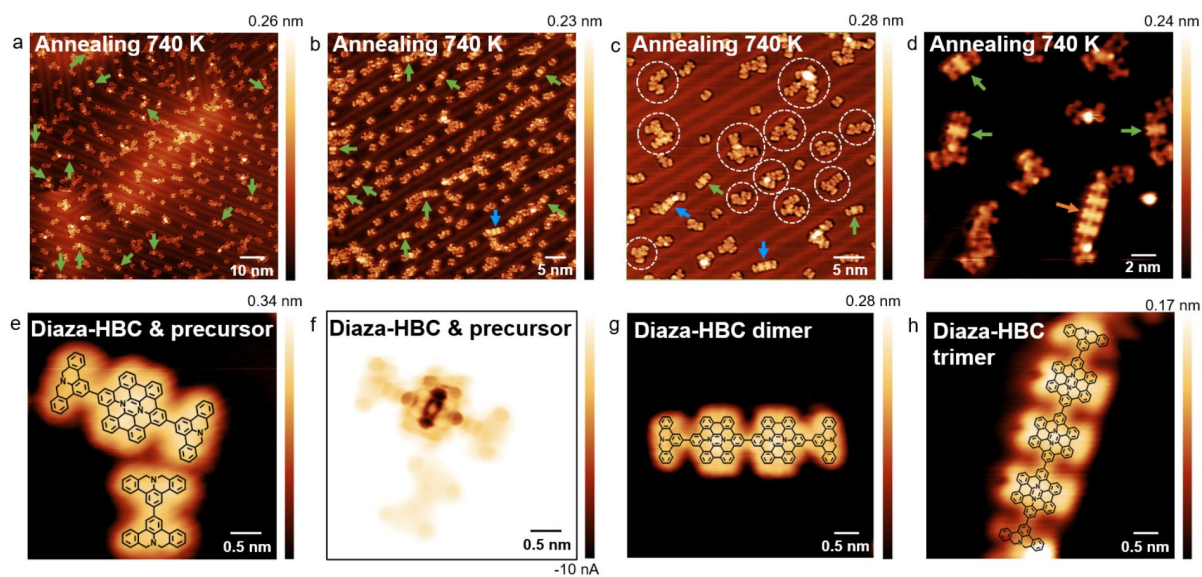

**Figure S2. a-d.** STM results of sample after 740 K annealing, polymers of different lengths are marked with arrows of different colors, and irregular clusters are marked with white dashed lines. a. $V_b$ = -150

mV, $I_t$ = 25 pA, T = 4.150 K. **b.** $V_b$ = -150 mV, $I_t$ = 25 pA, T = 4.488 K. **c.** $V_b$ = -200 mV, $I_t$ = 25 pA, T = 3.934 K. **d.** $V_b$ = -200 mV, $I_t$ = 25 pA, T = 4.475 K. **e-h.** Zoomed-in STM results after 740 K annealing, showing precursor, **3•⁺**, dimer, and trimer, respectively. **e.** $V_b$ = -50 mV, $I_t$ = 25 pA, T = 2.322 K. **f.** $V_b$ = -50 mV, T = 2.507 K. **g.** $V_b$ = -50 mV, $I_t$ = 25 pA, T = 3.497 K. **h.** $V_b$ = -200 mV, $I_t$ = 25 pA, T = 4.475 K.

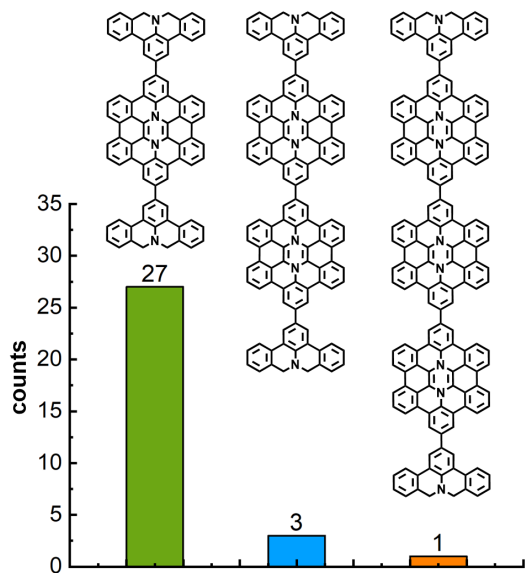

**Figure S3.** Histogram of the number of diPAMY-based molecular chains on a 110 × 110 nm Au(111) surface and their corresponding structures. The side group termination is not identified in this graph.

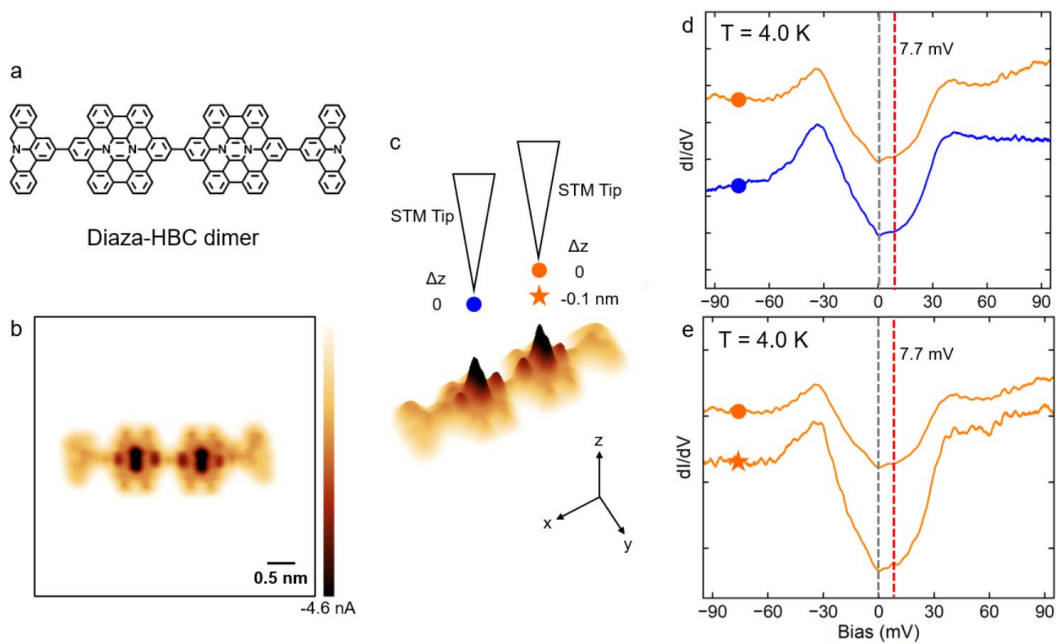

**Figure S4. a.** Structure of **4**. **b.** STM image, $V_b$ = -50 mV, T = 4.480 K. **c.** Topographic data depicting the tip position variation at d and e. $V_b$ = -50 mV, T = 4.480 K. **d.** Scanning tunneling spectroscopy of **4** at different positions. **e.** Height dependent scanning tunneling spectroscopy of **4** at 4.0 K.

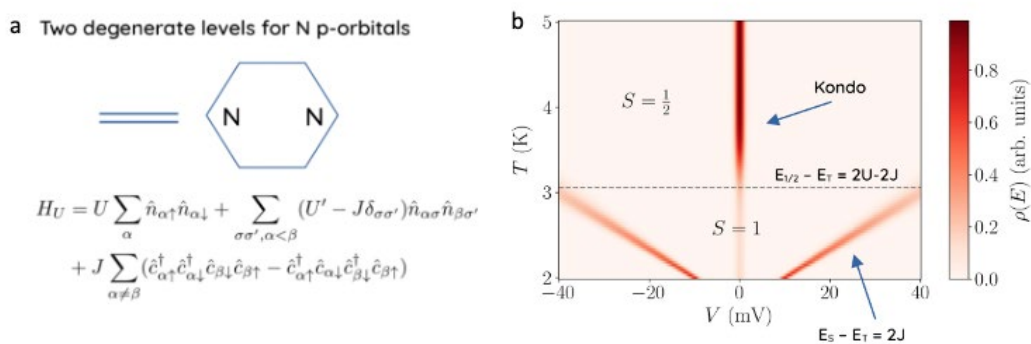

**Figure S5.** Alternative spin model with a triplet as a ground state originating from **3•** triplet state at lower temperatures. Aa two-orbital model supplemented with a strong local Coulomb interaction (a Hubbard U) as well as a temperature dependent exchange coupling (a Hund's J), Assuming that J depends linearly on temperature, a linearly dispersing singlet-triplet transition below $T_c$ ~ 3K is reproduced, as well as the transition from a triplet S = 1 ground state (for U > J) to a spin-1/2 ground state (for J > U). A system in the spin-1/2 regime it would lead to the single ZEP Kondo resonance. This model was not considered since the DFT calculations predict a doublet **3•$^{++}$** and a three-orbital model and not a triplet on the surface.

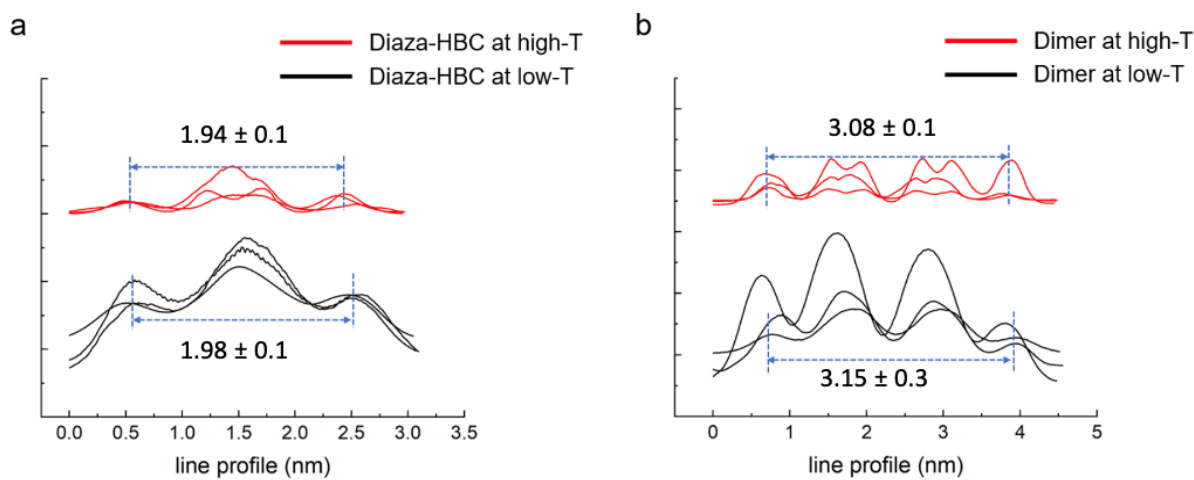

**Figure S6.** The line profiles of diaza-HBC **3** and its dimer **4** at low and high temperatures, respectively.